%% file: main.tex
\documentclass[manuscript,screen,nonacm]{acmart}
\AtBeginDocument{%
  }

\setcopyright{acmlicensed}
\copyrightyear{2025}
\acmYear{2025}
\acmDOI{XXXXXXX.XXXXXXX}
\acmISBN{978-1-4503-XXXX-X/2018/06}

\usepackage{hyperref}
\usepackage{booktabs}
\usepackage{csquotes}
\usepackage{enumerate}
\usepackage{subcaption}
\usepackage{multirow}
\usepackage{rotating}
\usepackage{siunitx}
\usepackage{pdfpages}
\usepackage{enumitem}
\usepackage{xspace}
\usepackage{tabularx}
\usepackage{float}
\usepackage{pgfplots}
\usepackage{pgfplotstable}
\usepgfplotslibrary{groupplots}
\usetikzlibrary{calc,matrix}
\usepackage{tcolorbox}
\usepackage{threeparttable} 
\usepackage{svg}

 \pgfplotsset{compat=1.18}

\setlist[enumerate,1]{label=(\arabic*)}
\setlist[enumerate,2]{label=(\theenumi.\arabic*)}
\setlist[enumerate,3]{label=(\theenumii.\arabic*)}

\setlist[enumerate,1]{label=(\arabic*), ref=\arabic*}
\setlist[enumerate,2]{label=(\theenumi.\arabic*), ref=\theenumi.\arabic*}
\setlist[enumerate,3]{label=(\theenumii.\arabic*), ref=\theenumii.\arabic*}

\begin{document}

\title[A Late-Breaking Study on How LLM As-A-Service Customizations Shape Trust and Usage Patterns]{Campus AI vs Commercial AI: A Late-Breaking Study on How LLM As-A-Service Customizations Shape Trust and Usage Patterns}


\author{Leon Hannig}
\orcid{0009-0007-9212-9037}
\affiliation{%
  \institution{University of Duisburg-Essen}
  \city{Duisburg}
  \country{Germany}
}
\email{leon.hannig@uni-due.de}

\author{Annika Bush}
\orcid{0000-0003-1475-4260}
\affiliation{%
	\institution{TU Dortmund University}
	\city{Dortmund}
	\country{Germany}
}
\email{annika.bush@tu-dortmund.de}

\author{Meltem Aksoy}
\orcid{0000-0003-3232-3923}
\affiliation{%
  \institution{TU Dortmund University}
  \city{Dortmund}
  \country{Germany}
}
\email{meltem.aksoy@tu-dortmund.de}

\author{Steffen Becker}
\orcid{0000-0001-7526-5597}
\affiliation{%
  \institution{Ruhr University Bochum}
  \city{Bochum}
  \country{Germany}
}
\additionalaffiliation{%
  \institution{Max Planck Institute for Security and Privacy}
  \city{Bochum}
  \country{Germany}
}
\email{steffen.becker@rub.de}

\author{Greta Ontrup}
\orcid{0000-0003-4720-1494}
\affiliation{%
  \institution{University of Duisburg-Essen}
  \city{Duisburg}
  \country{Germany}
}
\email{greta.ontrup@uni-due.de}

\renewcommand{\shortauthors}{Hannig et al.}

\input{section/00_abstract}

\input{keywords}

\maketitle

\input{section/01_introduction}

\input{section/02_background}

\input{section/03_methods}

\input{section/04_Conclusion}

\input{acknowledgements}

\bibliographystyle{ACM-Reference-Format}
\bibliography{bibliography}


\end{document}

%% file: section/00_abstract.tex
\begin{abstract}
As the use of Large Language Models (LLMs) by students, lecturers and researchers becomes more prevalent, universities -- like other organizations -- are pressed to develop coherent AI strategies. 
LLMs as-a-Service (LLMaaS) offer accessible pre-trained models, customizable to specific (business) needs.
While most studies prioritize data, model, or infrastructure adaptations (e.g., model fine-tuning), we focus on user-salient customizations, like interface changes and corporate branding, which we argue influence users' trust and usage patterns. 
This study serves as a functional prequel to a large-scale field study in which we examine how students and employees at a German university perceive and use their institution's customized LLMaaS compared to ChatGPT.
The goals of this prequel are to stimulate discussions on psychological effects of LLMaaS customizations and refine our research approach through feedback. 
Our forthcoming findings will deepen the understanding of trust dynamics in LLMs, providing practical guidance for organizations considering LLMaaS deployment.
\end{abstract}

%% file: keywords.tex
\ccsdesc[500]{Human-centered computing~Empirical studies in HCI}
\ccsdesc[300]{Human-centered computing~User studies}
\ccsdesc[300]{Human-centered computing~Field studies}
\ccsdesc[100]{Security and privacy~Privacy protections}

\keywords{Large Language Models as-a-Service (LLMaaS), Customization, Trust, Hallucinations, University}

%% file: section/01_introduction.tex
\section{Introduction}
\label{sec::introduction}

As Large Language Models (LLMs) become increasingly prevalent in professional and educational settings, organizations face mounting pressure to develop coherent strategies for AI integration \cite{kapania2024}. 
Current research shows high adoption rates of commercial off-the-shelf LLMs such as ChatGPT among students and university staff for various professional and academic tasks \cite{Zhang2024, Zhou2023, vonGarrel2023}. 
In order to guide the informed and secure use of AI, many organizations, including universities, corporations, and government agencies, are exploring the deployment of LLMs as-a-Service (LLMaaS) as an achievable and cost-effective way to adopt AI in-house \cite{laMalfa2023}. This trend is driven by several practical considerations: the opportunity to offer access to all university members while maintaining institutional oversight and alignment with organizational policies or the ability to provide dedicated technical support \cite{paavola2024}. 
One of the main benefits of AI as-a-Service (AIaaS) solutions is the flexibility they offer in customizing various aspects \cite{Diaferia2022}, such as fine-tuning models or displaying user interfaces in corporate design.

Customizations are important because they can, e.g., improve model performance for the specific organizational context or enhance security and privacy safeguards \cite{sundberg2023, Lewicki2023}. 
However, even choices that do not impact model performance or security, such as branding a user interface, could lead to significant changes in how the system is perceived and used.

From a psychological perspective, users rely on various cues to assess the trustworthiness and consequently align their use of AI \cite{Schlicker2023,deVisser2014}. 
Users interpret information about and from the system in terms of its performance, purpose, and processes \cite{deVisser2014}. 
We argue that end users may interpret customizations of LLMs as indicative of the system's capabilities, goals, and underlying processes, even if the design choices (e.g., corporate branding) do not directly impact functionality.

While organizations invest in LLMaaS, we lack systematic empirical evidence on how customization choices within the adoption process influence users' trust, usage patterns, and overall acceptance of these systems compared to commercial off-the-shelf alternatives (hereinafter referred to as \enquote{commercial LLMs}). 
This is especially relevant for customization choices that are salient to the user (e.g., user interface) but have no real technical impact on the pre-trained data/model/algorithm/infrastructure \cite{Diaferia2022}. 
We aim to close this research gap by conducting an empirical study that answers the following research question:

\fcolorbox{green!75!black}{green!5!white}{\parbox{0.95\linewidth}{\textbf{RQ:} How do customizations of LLMaaS impact end users' trust, perception, and usage of LLMs in professional and academic contexts compared to commercial AI alternatives?}}

This late-breaking work presents the research motivation, hypotheses, and study design of an ongoing large-scale field study. The purpose of this \enquote{prequel} is twofold: first, we compile a comprehensive overview of LLMaaS customization options to guide the hypothesis formulation, aiming to stimulate further discussion and research into the psychological effects of customization choices on end users. 
Second, we seek feedback from our professional peers on our proposed field study to enhance its effectiveness. 
Outcomes from our large-scale field study will enrich both theoretical insights into trust in organizational AI systems and practical applications of AI strategies across different sectors.

%% file: section/02_background.tex
\section{Background}
\label{sec::background}

\subsection{LLMs (as-a-Service) at Universities}
\label{subsec::background::LLMs_at_Universities}

LLMs, such as ChatGPT, have become increasingly prevalent in academic settings, with significant adoption among university students. 
For example, \citet{vonGarrel2023} report that nearly two thirds of German students use LLMs for tasks that include concept clarification, literature review, and translation. 
Similarly, \citet{amani2023generative} demonstrate the widespread use of LLMs for personalized learning purposes at Texas A\&M University. 
Researchers have reported using LLMs \cite{kapania2024, syed2024awareness}, e.g., for language editing or literature search \cite{abdelhafiz2024knowledge}. 

Thus, universities are caught between enabling students and staff to access and use AI in an informed way to leverage the benefits while also protecting data, privacy, rights, and ethical principles. 
One way to appropriately shape the use of AI is for universities to introduce \enquote{their own} LLMs, the same way as organizations in the private sector do (e.g., WärtsiläGPT by the Finnish company Wärtsilä Oyi; \cite{paavola2024}). 
However, the introduction of AI poses a massive challenge for universities, due to a lack of internal technical abilities and knowledge about the deployment of AI and the high costs of developing and maintaining a sufficient IT infrastructure \cite{lins2021}.

LLMs as-a-Service (LLMaaS) provide a practical solution by offering easy access to pre-trained models via cloud providers \cite{wang2024}. 
This approach is a subset of AI as-a-Service (AIaaS), broadly defined as \enquote{cloud-based systems providing on-demand services to organizations and individuals for deploying, developing, training, and managing AI models} \citep[p. 424]{lins2021}. 
Often referred to as \enquote{no-code or low-code} AI, these systems are recognized as a means of democratizing AI, enabling organizations to adopt and afford AI technologies without requiring extensive technological expertise \cite{sundberg2023}. 
Moreover, LLMaaS solutions emphasize enhanced data protection and security, offering a viable option for institutions like universities to leverage AI's advantages while adhering to stringent compliance and security standards \cite{paavola2024}. 
The primary benefits of AIaaS include reducing complexity, increasing automation, and enabling customization~\cite{lins2021}.

One proposed advantage of LLMaaS is that organizations can tailor various aspects of LLMs to better align with their unique requirements and objectives. In this context, customizations are defined as configurations of parameters at various architectural layers to improve the fit between the system and organization-specific needs \cite{Diaferia2022}. \autoref{tab:categories_features} outlines customization options, which encompass customizations of data, models, algorithms, and infrastructure \cite{Diaferia2022}, security and governance adaptations \cite{Chen2024}, customizations that relate to the design of the user interface and experience, and choices regarding the organizational integration. 

\begin{table}[htb]
    \centering
    \caption{Examples of Customization Aspects of AIaaS.}
    \label{tab:categories_features}
    \renewcommand{\arraystretch}{1.3}
    \begin{tabular}{p{3cm}p{9cm}p{1.8cm}}
        \toprule
        \textbf{Category} & \textbf{Examples} & \textbf{References} \\
        \midrule
        Model fine-tuning and Configuration & 
        Adjust/add system prompts, fine-tuning on proprietary data, hyperparameter configuration, domain-specific knowledge integration, customizable output. & \cite{lins2021, Diaferia2022, laMalfa2023, Hajipour2023, Parthasarathy2024}\\
        \midrule
        Integration and Interoperability & 
        API integration, integration with other tools and systems. & \cite{gan2023, lins2021}\\
        \midrule
        Security and Governance & 
        Access control and security, compliance and regulatory requirements, monitoring and evaluation. & \cite{gan2023, Doosthosseini2024, Parthasarathy2024, Iqbal2024}\\        
        \midrule
        User Interface and Experience & 
        Incorporation of corporate design, sustainability measures (e.g., token visualization), user profiling and personalization, customizable error handling, multi-language support. & \cite{Casati2019}\\
        \midrule
        Organizational integration & 
        Providing context-specific training or workshops, providing context-specific information and use recommendations regarding the model. & \cite{nagarajan2024}\\
        
        \bottomrule
    \end{tabular}
    \vspace*{\fill} 
\end{table}

To date, existing literature has focused primarily on methods and effects of model fine-tuning and configuration, integration and interoperability, as well as security and governance-related consequences of LLMaaS. 
These aspects are critical, as adapting AI systems -- for example, to ensure data protection or to fine-tune them for internal company data -- plays a key role in their secure and human-centered integration into organizations. 
However, other customization choices, such as interface design, remain largely unexplored in terms of their impact on end users. For instance, research and reports on LLMaaS deployment in organizations focus mostly on technical customizations, mentioning other customizations pertaining to the user interface and experience or the organizational integration as a side note (e.g., \citet{Weber2024} report on the architectural designs of Frauenhofers LLMaaS \enquote{FhGenie}). 
This could be because these adjustments have no effect on the actual performance or security of the models. 
However, from a psychological perspective, users' trust and corresponding use of AI depend on so-called cues, i.e. \enquote{any information element that can be used to make a trust assessment about an agent} \citep[p. 253]{deVisser2014}.
Users have been shown to rely on various cues when deciding whether or not to trust and use AI, e.g., system explanations or a logo of a company \cite{Schlicker2023}. \citet{gong2024} for example theorize on the effects of user experience (UX) design on perceived trustworthiness.
We therefore posit that it is essential to recognize that end users could perceive customizations of LLMs as reflections of the system's capabilities, goals, and inner workings, even if these customizations (such as using a corporate design) do not actually affect the functioning of the system. 

This study investigates the usage patterns and trust perceptions of end users when interacting with commercial  LLMs and customized LLMaaS.
In the following, we derive hypotheses based on previous research to assess how the customization choices of LLMaaS affect the perception and use of the system by users compared to a commercial LLM.

\subsection{Trust in AI Systems}

Trust, as a key determinant of technology adoption \cite{van2019trust, Choudhury2023}, has been recognized as an important factor for LLM adoption among academic staff and students \cite{bhaskar2024shall, Tiwari2024}. Research has shown that both insufficient and excessive trust in AI systems can be critical, as a lack of trust may hinder their adoption, while overtrust can lead to risks such as reliance on incorrect or potentially harmful outputs \cite{Bucinca2021, wagner2018overtrust}. 
This underscores the importance of ensuring that students and staff place appropriate trust in LLM chatbots. 

Trust is commonly defined as the trustor's positive expectations towards the trustee, combined with the trustor's vulnerability and uncertainty within the specific context \cite{balayn2024, lee2004trust}. 
Research has shown that trust levels in human-human interaction depend on three perceived qualities of the trustee: ability, integrity, and benevolence \cite{mayer1995}. 
In the context of human-automation interaction, these three factors have been reframed as performance, process (e.g. availability, confidentiality, understandability of a system) and purpose (e.g., the developer's intentions) \cite{lee2004trust}. 
Additional factors have been identified as particularly important for trust in LLMs. 
These include transparency \cite{huang2023, schwartz2023}, access to a human operator \cite{nordheim2019}, and perceived privacy and robust data security \cite{huang2023, pawlowska2024}. 
These insights provide a robust basis for examining how users develop trust when choosing between organizations' customized LLMaaS and commercial LLMs.

Customizations of LLMaaS in corporate design could serve as a salient trust cue for users, increasing the perceived trustworthiness of the system in two dimensions: (a) purpose and (b) process. 
The perceived purpose of the system may appear more trustworthy when the customizations incorporate university branding, as the university is likely to be seen as a benevolent provider with minimal financial or marketing-driven motives \cite{nordheim2019}. 
This perception may be further enhanced due to a sense of familiarity evoked by the university's branding \cite{Alarcon2016}. 
Second, the process dimension of the system may be perceived as more trustworthy because it is easily accessible and embedded in the existing organizational framework. 
Here, features such as the option to contact a human operator for assistance could contribute to users' trust through perceived openness and accessibility \cite{lee2004trust}. 
Additionally, a customized LLMaaS may be perceived as more transparent due to context-specific information about the system available on the university's website. 
These factors together lead us to the following hypothesis: 

\fcolorbox{blue!75!black}{blue!5!white}{\parbox{0.95\linewidth}{\textbf{H1:} Users report higher levels of trust in a customized LLMaaS compared to a commercial LLM.}}

We further hypothesize that higher trust in the organization's customized LLMaaS will likely depend on its members' trust in the organization itself. 
Organizational trust describes the extent to which individuals trust an organization~\cite{Pirson2011}. 
Prior research on AI implementation \cite{Lapinska2021} and attitudes towards LLMs \cite{balayn2024} suggests that organizational trust can significantly shape users' trust in the system. 
In line with this, \citet{nordheim2019} have demonstrated that customers' trust in a company's service chatbot is influenced by their trust in the company itself. 
We therefore assume: 

\fcolorbox{blue!75!black}{blue!5!white}{\parbox{0.95\linewidth}{\textbf{H2:} The level of organizational trust moderates the effect of system type (customized vs. commercial) on user trust, such that the effect is stronger when organizational trust is higher and weaker when it is lower.}}

\subsection{LLM Hallucinations} 

When implementing LLMaaS, universities face the challenge of ensuring that their students and staff can interact with these systems effectively, correctly and safely. 
A key aspect of this responsibility is addressing the issue of hallucinations. 
Hallucinations in LLMs describe the generation of content that appears plausible but is either factually incorrect or inconsistent with user input \cite{Huang_2024}. 
Hallucinations in human-AI interaction pose a dual challenge: they impair effective collaboration through reduced output quality and undermine users' trust \cite{gong2024, oelschlager2024}. 
In the academic context, hallucinations present particularly serious risks. 
For example, LLMs have been shown to often fabricate inaccurate scientific references~\cite{Aljamaan2024, Chelli.2024}. 
For university staff and students, hallucinations could thus lead to severe consequences. 

Despite various efforts and technological advances, hallucinations cannot be completely eliminated from LLMs~\cite{xu2024,banerjee2024}. 
This issue calls for a focus on the human side, with the ultimate goal of identifying strategies to help users anticipate, mitigate, and react to hallucinations. 
In this context, we consider two customization aspects of the user interface: corporate design and warnings about false information. 
Research has shown that the presence of logos from reputable companies can create positive associations, enhancing the perceived credibility of mobile systems \cite{lowry2005}. 
We argue that branding a chatbot with a university logo leads users to interpret the chatbot as university-approved, i.e., reliable and infallible. 
This could potentially make users less critical and cautious in their interactions with the university-branded system. 
Another notable customization choice of LLMaaS is the display of warnings about hallucinations or false information. 
Research suggests that such warnings can help users detect hallucinations more frequently \cite{nahar2024}. 
While ChatGPT's interface includes a disclaimer below the prompt box stating \enquote{ChatGPT can make mistakes. Please consider checking important information}, LLMaaS allow such warnings to be adapted, removed, or altered. 
We argue that making hallucination warnings less visible leads users to be less cautious about hallucinations in their LLM interactions, and thus to detect fewer hallucinations. 
Based on this reasoning, we propose the following hypotheses:

\fcolorbox{blue!75!black}{blue!5!white}{\parbox{0.95\linewidth}{\textbf{H3:} Users report less cautious behavior towards hallucinations when using a customized LLMaaS compared to a commercial LLM.}}

\fcolorbox{blue!75!black}{blue!5!white}{\parbox{0.95\linewidth}{\textbf{H4:} Users report fewer experienced hallucinations when using a customized LLMaaS compared to a commercial LLM.}}

\subsection{Data Security and Privacy Concerns}
End users often express concerns about data security and privacy when interacting with LLMs.
Key risks include (unauthorized) data retention~\cite{COBBE2021105573}, where providers may retain and use customer data to enhance their models, inference attacks that aim to extract sensitive information from trained models~\cite{kandpal-etal-2024-user}, and the unintended exposure of confidential data during model training or operation~\cite{chen2024janus}.
Technical customization options in LLMaaS offer promising avenues to mitigate these risks, thereby playing a pivotal role in enhancing users' perceptions of privacy and security while fostering trust.
These options include user-configurable privacy settings that empower users to control how their data is stored and used~\cite{10.1145/644527.644538}, transparency mechanisms like privacy explanations~\cite{BRUNOTTE2023111545}, which provide clarity about data handling practices, and the implementation of privacy-preserving techniques such as differential privacy~\cite{abadi2016deep} during training and fine-tuning processes. 
Beyond their direct technical benefits, we posit that customization measures, even those without tangible impacts on privacy or security, can still foster higher perceived privacy levels.
In this context, we again draw on the role of corporate branding as a salient customization feature: 
institution-branded LLMs could enhance data security and privacy perceptions, which in turn could lead users to perceive the system as more trustworthy, believing that their data will be handled with care and not shared with third parties. 
In this context, we propose the following hypothesis:

\fcolorbox{blue!75!black}{blue!5!white}{\parbox{0.95\linewidth}{\textbf{H5:} Users perceive greater privacy when using a customized LLMaaS compared to a commercial LLM.}}

\subsection{Sustainable AI Use}
\label{subsec::background::sustainable}

Having explored potential impacts of customization on users' trust, perceived hallucinations and privacy concerns, it becomes evident that organizations need to carefully consider their approach to LMMaaS. 
In this last part, we want to open the door for broader considerations that pertain to the global strategy of an organization, specifically the corporate social responsibility (CSR). 
CSR refers to situations in which organizations pursue strategies that go beyond business needs and legal regulations but aim at furthering social good \cite{McWilliams2001}. 
This includes various aspects, such as improving an organization's sustainability and environmental performance~\cite{McWilliams2006}.
We argue that customization of LLMaaS offer organizations opportunities to align AI deployment with their CSR strategy and corresponding sustainability goals, thereby not only shaping the trust in and perception of the systems but also their environmental impact.

The environmental impact of AI systems and LLMs has become an increasingly critical concern in computing research. 
Recent work by \citet{Rafael.2024} emphasizes how the invisibility of digital energy consumption often leads to unconscious and potentially wasteful usage patterns. 
This mirrors findings in energy consumption studies \cite{McCarthy.2023} that identify varying levels of user awareness -- from \enquote{energy unaware} to \enquote{energy aware and active} -- highlighting how visibility of resource consumption can transform user behavior. 
Our research extends these insights and considers the impact of customization choices aimed at promoting sustainable AI usage patterns.

Making resource consumption visible has shown promising results across different domains. \citet{Penkert.2023} demonstrated that transparency in sustainability information significantly influences user choices, while \citet{Sanduleac.2017} found that real-time energy data promotes more responsible consumption patterns. 
LLMaaS offer customization options to make AI resource limitations explicit through token visibility. 
We argue that customizing token visibility can foster what we term \enquote{AI resource consciousness} -- a more deliberate and efficient approach to AI interaction that considers both immediate utility and broader environmental impact.
Based on these previous findings, we hypothesize: 

\fcolorbox{blue!75!black}{blue!5!white}{\parbox{0.95\linewidth}{\textbf{H6:} Users report (a) more resource-efficient prompting behaviors and (b) increased sensitivity to AI sustainability concerns when using a customized LLMaaS compared to a commercial LLM.}}

%% file: section/03_methods.tex
\section{Method}

\subsection{Study Design, Participants and Setting}

To answer our hypotheses, we plan a quantitative, cross-sectional field study. 
We designed a comprehensive survey instrument combining validated measures from existing research with questions specific to the use of LLM-based chatbots. 
In the following, we report on the planned design, method and questionnaire.

The large-scale study will be conducted in the spring of 2025 at a large German research university that has implemented a customized LLMaaS in late 2024.
The system has approximately 6,000 registered users and 800 active users per week as of January 22, 2025.
Our participant pool includes three distinct user groups within the university ecosystem: (a) students across all academic levels and disciplines, (b) research staff, including PhDs, PostDocs, faculty, and (c) administrative staff. 
A sample size of $N = 250$ (accounting for dropout) is planned ($f^2 = 0.05$, $\alpha = 0.05$, $\beta = .80$). 
The goal is to ensure an equal distribution of participants between users of the customized LLMaaS and ChatGPT as well as an equal representation across the three status groups.

\subsection{Customized LLMaaS}
The LLM-based chatbot introduced by the university is OpenAI's ChatGPT deployed via Microsoft Azure. 
Customizations regarding the data and model were kept to a minimum, but the university made some adjustments regarding the user interface and incorporated detailed information and training materials (see \autoref{tab:customized_LLM}).

\begin{table}[htb]
    \centering
    \begin{threeparttable}
    \caption{Customization features of the studied LLMaaS.}
    \label{tab:customized_LLM}
    \begin{tabular}{p{4.5cm}p{9.5cm}}
        \toprule
        \textbf{Category} & \textbf{Customization Details} \\
        \midrule
        Model fine-tuning \& Configuration & 
        Temperature\tnote{1}\hspace{0.2em} set to 0 for deterministic outputs \\
        \midrule
        Integration and Interoperability & 
        Access to multiple models; image generation via DALL-E for OpenAI models \\
        \midrule
        Security and Governance & 
        Minimal data collection for service provision; chat data excluded from model training; GDPR-compliant data processing within the EU; access restricted to internal networks (e.g., via VPN)  \\        
        \midrule
        User Interface and Experience & 
        Interface customized with corporate design (e.g., university logo); token usage visualized as a percentage in the user interface \\
        \midrule
        Organizational integration & 
        Supporting materials provided on the chatbot landing page and intranet; training sessions on effectively using LLMs in higher education \\
        \bottomrule
    \end{tabular}
    \begin{tablenotes}
        \footnotesize
        \item[1] The temperature parameter in LLMs, ranging from 0 to 1, controls response randomness, with values near 0 favoring deterministic outputs and values near 1 producing diverse, creative outputs.
    \end{tablenotes}
    \end{threeparttable}
\end{table}

\subsection{Survey Instrument}
\begin{figure}[h]
\centering
\includegraphics[width=0.7\textwidth]{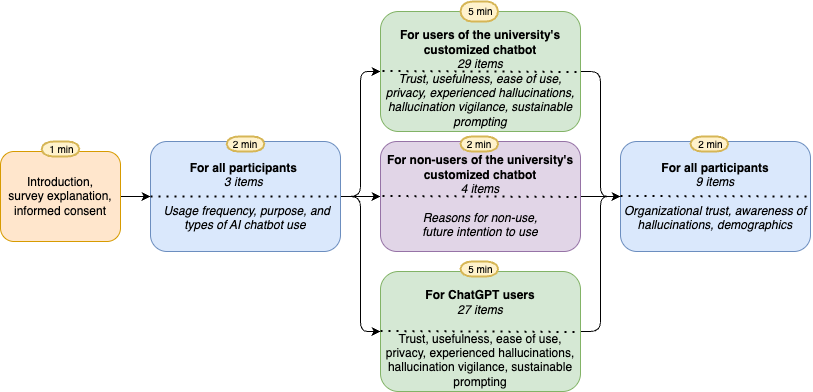}
\caption{Overview of our survey. The expected completion time is 15 minutes or less.}
\Description{}
\label{fig:Survey_Flow}
\end{figure}

We designed a comprehensive survey to investigate the perceptions and usage patterns of both the university's customized LLMaaS chatbot and OpenAI's ChatGPT. 
The proposed survey design is summarized in \autoref{fig:Survey_Flow}. 
Participants are guided through the survey based on their use/non-use of the university-customized chatbot and ChatGPT. 
The main part consists of questions regarding the perception and use of the university's system and ChatGPT (green blocks in \autoref{fig:Survey_Flow}). The survey questions are available in the supplementary material.

The survey begins by capturing usage patterns, including frequency of use (ranging from \enquote{never} to \enquote{multiple times per day}) and the context of use (private versus work/study-related). 
In the first step, several aspects of system perception and use are explored: the context of use (specific tasks), use of training/information material, perceived usefulness (productivity enhancement, task facilitation), and ease of use (interface clarity, learning curve), both assessed following the Technology Acceptance Model (TAM)\cite{davis1989}.
For users familiar with either system, we also assess their awareness and use of different language models available within each platform. 
For users who have never used or rarely use the university's chatbot, we explore their reasons for non-adoption or limited use.

The main part of the survey is structured around the key constructs (see hypotheses), measured in parallel for both systems. 
Participants who use both systems will respond to the same set of items twice: once for the university's customized LLMaaS and once for ChatGPT. 
At the beginning of the main part, participants are reminded to answer the following questions specifically in regard to their work-related or academic use of the chatbots, and not for their private use. 
We use five-point Likert scales ranging from \enquote{strongly disagree} to \enquote{fully agree}. 

\paragraph{Trust} Trust is measured along two dimensions: trust in the providing organization (university) is measured with three items following \citet{Boe2018} (e.g., \enquote{My university will safeguard my interests.}) and trust in the chatbots themselves is assessed by the facets of global trust, benevolence/purpose, integrity/process and ability/performance using four items from \citet{wischnewski2023trustAI} (e.g., \enquote{I trust [the university's chatbot/ChatGPT].}). 

\paragraph{Hallucinations} We investigate users' cautious behavior towards LLM hallucinations, including their verification practices and task-specific trust decisions, with three items (e.g., \enquote{I check important information from [the university's chatbot/ChatGPT] responses through external sources.}). Users also report their experienced hallucinations with three items (e.g., \enquote{I have noticed that [the university's chatbot/ChatGPT] makes up facts that do not correspond to reality.}). The items are self-developed based on \citet{Christensen2024}.

\paragraph{Data security and privacy concerns} We measure privacy concerns following \citet{Hsu2018} with three items (e.g., \enquote{My decision to use [the university's chatbot/ChatGPT] exposes me to privacy risks.}).

\paragraph{Sustainable AI use} The survey also assesses the intentions for sustainable use, environmentally conscious behavior, and considerations of energy consumption with three items (e.g., \enquote{I often think about the energy consumption of [the university's chatbot/ChatGPT].}). Since the university-chatbot displays the remaining number of tokens as a percentage, we assessed the university-chatbot users' awareness of token usage with two items (e.g., \enquote{During use, I pay attention to how many tokens I still have available.}). 
Because there are no tested surveys focusing explicitly on the sustainable use of LLMs yet, the items are self-developed.

\subsection{Data Collection Procedure}
To maximize participation and ensure representative sampling, we employ a multi-channel recruitment strategy through the university's chatbot landing page, department office communications, university-wide communication channels, and supplementary physical materials where needed.
The university communications department will assist in disseminating the survey through official social media channels and website postings.

%% file: section/04_conclusion.tex
\section{Outlook: Expected Contributions}

Our research provides three key contributions to the field: 
First, for higher education institutions and organizations in general, we offer evidence-based insights for AI implementation strategies, helping decision-makers understand how customization choices of LLMaaS solutions impact user trust and adoption patterns. 
Second, we develop design recommendations for organizational AI systems, addressing how features, interfaces, and integration approaches can be optimized to enhance user trust and adoption while effectively coexisting with commercial alternatives. 
Third, we open new research directions for investigating how customization aspects, especially organizational AI branding effects, vary across sectors, examining long-term impacts on organizational culture, and exploring the evolving relationship between institutional trust and AI system trust over time.

%% file: acknowledgements.tex
\begin{acks}
This work was supported by the \grantsponsor{rctrust}{Research Center Trustworthy Data Science and Security}{https://rc-trust.ai} (\url{https://rc-trust.ai}), one of the Research Alliance Centers within the UA Ruhr (\url{https://uaruhr.de}).
\end{acks}